\magnification=\magstep1
\tolerance=500
\bigskip
\rightline{25 November, 2012}
\bigskip
\centerline{\bf Spin, angular momentum and spin-statistics for a}
\centerline{{\bf relativistic quantum many body
system}\footnote{*}{This paper is dedicated to the memory of Constantin Piron.}}
\bigskip
\centerline{Lawrence Horwitz}
\smallskip
\centerline{School of Physics, Tel Aviv University, Ramat Aviv
69978, Israel}
\centerline{Department of Physics, Bar Ilan University, Ramat Gan
52900, Israel}
\centerline{Department of Physics, Ariel University Center of
Samaria, , Ariel 40799, Israel} 

\bigskip
\noindent PACS:03.30.+p, 03.65 Pm, 03.65.-w, 03.65.Ud, 12.20.-m,
03.70.+k, 11.10.-z
\smallskip
\noindent Key words: Relativity, spin, induced representation,
spin-statistics, quantum fields, fibration.

\bigskip
\noindent{\it Abstract} The adaptation of Wigner's induced
representation for a relativistic quantum theory making possible the
construction of wavepackets and admitting covariant expectation values
for the coordinate operator $x^\mu$ introduces a foliation on the
Hilbert space of states. The spin-statistics relation for fermions and
bosons implies the universality of the parametrization of orbits of the induced
representation, implying that all particles within identical particle
sets transform under the same $SU(2)$ subgroup of the Lorentz group,
and therefore their spins and angular momentum states can be computed
using the usual Clebsch-Gordan coefficients associated with angular
momentum. Important consequences, such as entanglement for subsystems
at unequal times, covariant statistical correlations in many body systems,
and the construction of relativistic boson and fermion statistical ensembles,
 as well as implications for the foliation of the Fock
space and for quantum field theory are briefly discussed.
\bigskip
\noindent{\bf 1. Introduction}
\bigskip
\par  The spin of a particle in a nonrelativistic framework
corresponds to the 
lowest dimensional nontrivial representation of  the rotation 
group; the generators are the Pauli matrices $\sigma_i$ divided by
two, the generators of the fundamental representation of
the double 
covering of $SO(3)$. The self-adjoint operators that are the
generators of this group measure intrinsic angular momentum and are
associated with magnetic moments. 
\par In the nonrelativistic quantum theory, the spin states of a two
or more particle system  are defined by combining the spins of these
particles at equal time using appropriate Clebsch-Gordan coefficients
[1] at each value of the time.  The restriction to equal time
follows from the tensor product form of the representation of the
quantum states for a many body problem [2]. This correlation at equal
time in the nonrelativistic quantum theory
  is the source of the
famous Einstein-Podolsky-Rosen discussion[3] and provides an
important model for quantum information transfer.
\par The standard Pauli description of a particle with spin is not, however,
relativistically covariant, 
but Wigner [4] has shown how to describe this dynamical property of
a particle in a covariant way. The method developed
by Wigner has
provided the foundation for what is now known as the theory of induced 
representations [5], with very wide applications, including a
very powerful approach to finding the representations of noncompact
groups.
\par The formulation of Wigner[4] is, however, not appropriate for
application to quantum theory, since it does not preserve, as we shall
explain below, the covariance of the expectation value of coordinate
operators. We first briefly review Wigner's
method in its original form, and show how the difficulties arise. We
then review the extension of Wigner's approach necessary to describe the spin  
of a particle in the framework of the manifestly covariant theory of
Stueckelberg, Horwitz and Piron [6](SHP). We then show
that the observed correlation of spin and statistics for identical
particles necessitates a structure for which the Hilbert space of
states of a many body system of identical particles is represented as
a direct integral over all values of a
(normalized) timelike vector, a structure called {\it foliation}.
The relativistic many
body system then admits the desription of total spin (in general,
total angular momentum) states through the computation of
Clebsch-Gordan coefficients as in the nonrelativistic case, and
implies correlations between the spins of the particles much in the
same way. It has been shown that relativistically covariant canonical
 ensembles can be constructed in the framework of the SHP
theory[7] as well as a corresponding Boltzmann
transport theory[8].  The results that we
achieve here admit an extension of these results to particles with
spin; the results obtained in these earlier works may be
embedded in the foliation implied by the accommodation of spin.  The
full development of the consequences for thermodynamics
and phase transitions will be left for succeeding studies.    
\par  The foliation universally induced in the representation for
physical many body systems applies both to fermion and boson sectors of
the full Fock space, and therefore to the quantum fields. Further
development of the consequences of this structure will also be left
for succeeding publications,
\par As Wigner [4] has shown (see also the detailed discussion in
Weinberg [9]), constructing a representation of the Lorentz group by
inducing a representation on the stability group of the (timelike) 
four-momentum, one obtains a representation $\psi(p,\sigma)$ with the
transformation property
$$ \psi'(p,\sigma) = \psi(\Lambda^{-1}p, \sigma')D_{\sigma',
\sigma}(\Lambda,p), \eqno(1)$$
under the action of the Lorentz group, taking into account the spin
degrees of freedom of the wavefunction, where the matrix
transformation factor (Wigner's ``little group''[4])is constructed of 
the $2 \times 2$ matrices of $SL(2,C)$.
\par  The presence of the $p$-dependent
matrices representing the spin of a relativistic particle in the
transformation law of the wave function, however,  destroys the
covariance, in a
relativistic quantum theory, of the expectation value of the 
coordinate operators [11] in states transforming as in $(1)$.  To see
this, consider the expectation value of
the dynamical variable $x^\mu$ [6], {\it i.e.}
$$ <x^\mu> = \Sigma_\sigma\int d^4 p \psi(p,\sigma)^\dagger i{ \partial 
\over \partial p_\mu} \psi(p,\sigma). \eqno(2)$$ 
\par A Lorentz transformation would introduce the $p$-dependent
$2\times 2$ 
unitary transformation  on the function $\psi(p,\sigma)$, and the derivative
with  respect to momentum would destroy the covariance property that
we would wish to see of the expectation value
$<x^\mu>$.  
\par It is also not possible, in this framework, to form wave
packets of definite spin
by ($4D$) Fourier transform over the momentum variable, since this would add 
functions over different parts of the orbit, with a different $SU(2)$
at each point.
\par These problems were solved [11] by inducing a representation of
the spin 
on a timelike unit vector, say, $n^\mu$ in place of the four-momentum. 
\par  Using a representation induced on a timelike vector $n^\mu$,
which is independent of $x^\mu$ or $p^\mu$ \footnote{**}{Note that the 
resulting Stueckelberg type wave functions 
$\psi_n(x,\sigma)$ are local [6] and do not have the non-local properties
discussed by Newton and Wigner [10].}
permits the linear
superposition of momentum eigenstates to
form wave packets of definite spin, and admits the construction of
definite spin states for many body relativistic systems.  In the
following, we show how such a
representation can be constructed, and discuss some of its dynamical
implications.
\bigskip
\noindent{\bf 2. Induced representation on timelike vector $n^\mu$}
\bigskip

\par  We briefly review here the construction given in [11] in order
to make clear the nature of the resulting foliation of the Hilbert
space. Let us define,
$$ |n,\sigma, x> \equiv U(L(n)) |n_0, \sigma, x>,
\eqno(3)$$
where we may admit a dependence on $x$ (or, through Fourier transform,
on $p$). Here, we distinguish the action of $U(L(n))$ from the general
Lorentz transformation $U(\Lambda)$;  $U(L(n))$ acts only on the
manifold of $\{n^\mu\}$.  Its infinitesimal generators are given
by
$$ M^{\mu\nu}_n = -i (n^\mu {\partial \over \partial n_\nu} - n^\nu 
{\partial \over \partial n_\mu}),
\eqno(4)$$
while the generators of the transformations $U(\Lambda)$ act on the
full space of both the $n^\mu$ and the $x^\mu$ (as well as
$p^\mu$); its generators are given by  
$$M^{\mu\nu} = M^{\mu \nu}_n + (x^\mu p^\nu - x^\nu p^\mu). \eqno(5)$$
The two terms of the generator commute, and therefore the full
group is a (diagonal) direct product.
\par We now investigate the properties
of a total Lorentz transformation, {\it i.e.}, as in Wigner's procedure[4],
$$ U(\Lambda^{-1}) |n,\sigma, x> = U(L(\Lambda^{-1}n)
 \bigl(U^{-1}(L(\Lambda^{-1}n))U(\Lambda^{-1})\bigr)
|n_0, \sigma, x>, \eqno(6)$$
Now, consider the conjugate of $(6)$, 
$$ <n,\sigma, x|U(\Lambda)= <n_0, \sigma,
x|\bigl(U(L^{-1}(n))U(\Lambda)U(L(\Lambda^{-1}n))\bigr)
U^{-1}(L(\Lambda^{-1}n)). \eqno(7) $$
\par  The operator in the first factor (in parentheses) preserves $n_0$, and 
therefore corresponds to an element of the little group associated
with $n^\mu$ which may be represented by the matrices of $SL(2,C)$. 
It also, due to the factor  $U(\Lambda)$( for which the
generators are those of the Lorentz group acting both on $n$ and $x$
(or $p$)), takes $x
\rightarrow \Lambda^{-1} x $ in the conjugate ket on the left. Taking
the product on both sides with $|\psi>$, we obtain
$$ <n,\sigma, x|U(\Lambda)| \psi> = <\Lambda^{-1}n, \sigma', \Lambda^{-1}x|\psi>
D_{\sigma', \sigma}(\Lambda, n), \eqno(8)$$
or [11]
$$ \psi'_{n,\sigma}(x)= \psi_{\Lambda^{-1}n, \sigma'}(\Lambda^{-1} x)
D_{\sigma', \sigma}(\Lambda, n).
\eqno(9)$$
 where 
$$ D(\Lambda, n) = L^{-1}(n)\Lambda L(\Lambda^{-1}n), \eqno(10)$$
with $\Lambda$ and $L(n)$ the corresponding
$2\times 2$ matrices of $SL(2,C)$.
\par It is clear that, with this transformation law, one may take the
Fourier transform to obtain the wave function in momentum space, and 
conversely.  The matrix $D(\Lambda,n)$ is an element of $SU(2)$, and therefore
linear superpositions over momenta or coordinates
maintain the definition of the particle spin for each $n^\mu$, and interference
phenomena for relativistic particles with spin may be studied
consistently. Furthermore, if two
or more particles with spin are represented in representations induced
on $n^\mu$, at the same value of $n^\mu$ on their respective orbits,
and therefore in the {\it same} $SU(2)$ representation,
their spins can be added by the standard methods with the use of
Clebsch-Gordan coefficients. This method therefore admits the
treatment of a many body relativistic system with spin, as in the
proposed experiment of Palacios, {\it et al} [12]. 
\par Our assertion of the unitarity of the $n$-dependent part
of the transformation has assumed that the integral measure on the
Hilbert space, to admit integration by parts,  is of the form $d^4 n
d^4 x $, where the support of the wave functions on $n^\mu$ is in the
timelike sector. The action of the generator of Lorentz
transformations on $n^\mu$ maintains the normalization of
$n^\mu n_\mu$, which we shall take to be $-1$ in our discussion of the
Dirac representation for the wave function. Although the timelike vector
$n^\mu$, in many applications, is degenerate, it carries a
probability interpretation under the norm, and may play a dynamical
role (for example, as for the spacelike
inducing vector for the two body bound state problem in the covariant
Zeeman formulation of ref.[13]).
\par  There are two
fundamental representations of $SL(2,C)$ which are
inequivalent[14]. Multiplication of a two dimensional spinor
representing one of these by the operator $\sigma\cdot p$, expected
to occur in any dynamical theory, results in
an object transforming like the
other representation, and therefore the state of lowest dimension in
spinor indices of a  physical system  should contain both
representations. As we shall emphasize, however,  in our treatment of the more
than one particle system, for the rotation subgroup, both of the
fundamental representations yield the same $SU(2)$ matrices up to a
unitary transformation, and therefore the Clebsch-Gordan decomposition
of the product state into irreducible representations may be carried
out independently of which fundamental $SL(2,C)$ representation is
associated with each of the particles. This is therefore true for the
Dirac representation, incorporating both fundamental representations,
as we see in Eq. $(11)$.
\par As in ref. [11], one finds the Dirac spinor 
 $$ \psi_n(x) = { 1 \over \sqrt{ 2}} \left(\matrix{1&1\cr -1 &
1\cr}\right) \left(\matrix{L(n) {\hat \psi}_n(x) \cr {\underline L}(n)
{\hat \phi}_n(x) \cr}\right), \eqno(11)$$
 which transforms as 
$$ \psi'_n (x) = S(\Lambda) \psi_{\Lambda^{-1}n} (\Lambda^{-1} x)
\eqno(12)$$
where $S(\Lambda)$ is a (nonunitary) transformation generated
infinitesimally, as in the standard Dirac theory (see, for example,
Bjorken and Drell[16]), by $\Sigma^{\mu\nu} \equiv { i \over 4}
[\gamma^\mu, \gamma^\nu]$.
\par Following the arguments of [11], one can construct, in the
presence of a $U(1)$ gauge field, the covariant Hamiltonian 
 
$$K = {1 \over 2M}(p-eA)^2 + {e \over 2M} \Sigma_n^{\mu\nu}
F_{\mu\nu}(x) - eA_5, \eqno(13)$$
where 
$$ \Sigma_n^{\mu\nu}= \Sigma^{\mu\nu} + K^\mu n^\nu -K^\nu
n^\mu, \eqno(14)$$
and $ K^\mu = \Sigma^{\mu\nu}n_\nu$. 
The $A_5$ field arises as a compensation field for the $\tau$
derivative in the Stueckelberg-Schr\"odinger equation [17]. In general,
in this framework, the $A^\mu$ and $A^5$
fields may depend on $\tau$, since they correspond to gauge
compensation fields for the local gauge transformation
$\psi_\tau(x) \rightarrow \exp{i\Lambda(x,\tau)}\psi_\tau(x)$. The
$\tau$-independent Maxwell fields correspond to the zero mode of the
$A^\mu$ fields used here [17]. The currents constructed from the
Lagrangian associated with $(13)$ are, according to $(11)$, also
foliated, and therefore the fields $A_\mu, A_5$ generated by these
currents, will be foliated as well.  We shall discuss this structure
further in a succeeding article.  
\par The expression  $(13)$ is quite similar to that of the second order Dirac
operator; it is, however, Hermitian and has no direct electric
coupling to the electromagnetic field in the special frame for which
$n^\mu = (1,0,0,0)$ in the minimal coupling model we have given here
(note that in his calculation of the anomalous magnetic moment[18],
Schwinger puts the electric field to zero; a non-zero electric field
would lead to a non-Hermitian term in the standard Dirac propagator,
the inverse of the Klein-Gordon square of the interacting Dirac
equation). Note that in the derivation of the anomalous magnetic
moment given by Bennett[19], this restriction is not necessary since
the generator of the interacting motion is intrinsically Hermitian.
\par The matrices
$\Sigma_n^{\mu\nu}$ are, in fact, a relativistically covariant form
of the Pauli matrices.
\par To see this[11], we note that the quantities $K^\mu$ and
$\Sigma_n^{\mu \nu}$ satisfy the commutation relations 
 $$\eqalign{ [K^\mu,K^\nu] &= -i \Sigma_n^{\mu\nu}\cr
[\Sigma_n^{\mu\nu}, K^\lambda] &= -i[(g^{\nu\lambda} + n^\nu n^\lambda)
K^\mu - (g^{\mu\lambda} + n^\mu n^\lambda) K^\nu, \cr
[\Sigma_n^{\mu\nu}, \Sigma_n^{\lambda\sigma}] &= -i[(g^{\nu\lambda} +
n^\nu n^\lambda)\Sigma_n^{\mu\sigma} -(g^{\sigma\mu} + n^\sigma n^\mu)
\Sigma_n^{\lambda\nu} \cr
&-(g^{\mu\lambda} + n^\mu n^\lambda)\Sigma_n^{\nu\sigma} +
(g^{\sigma\nu} + n^\sigma n^\nu) \Sigma_n^{\lambda\nu}].\cr}
\eqno(15)$$
Since $K^\mu n_\mu = n_\mu\Sigma^{\mu\nu} = 0$, there are only three
independent $K^\mu$ and three $\Sigma_n^{\mu\nu}$. The matrices 
$\Sigma_n^{\mu\nu}$ are a covariant form of the Pauli matrices, and the last of
$(15)$ is the Lie algebra of $SU(2)$ in the spacelike surface
orthogonal to $n^\mu$. The three independent $K^\mu$ correspond to
the non-compact part of the algebra which, along with the
$\Sigma_n^{\mu\nu}$ provide a representation of the Lie algebra of the
full Lorentz group.  The covariance of this representation follows from
$$ S^{-1} (\Lambda) \Sigma_{\Lambda n}^{\mu\nu}S(\Lambda)
\Lambda_\mu^\lambda \Lambda_\nu^\sigma = \Sigma_n^{\lambda\sigma} . 
\eqno(16)$$
\par In the special frame for which  $n^\mu = (1,0,0,0))$,
$\Sigma_n^{ij}$ become the Pauli matrices ${1\over 2} \sigma^k$
with $(i,j,k)$ cyclic, and $\Sigma_n^{0j} = 0$. In this frame there is
no direct electric interaction with the spin in the minimal coupling
model $(14)$.  We remark that there is, however, a natural spin
coupling which becomes pure electric in the special frame[11],
given by (in gauge covariant form)
$$ i[K_T,K_L] = -ie \gamma^5 (K^\mu n^\nu - K^\nu n^\mu) F_{\mu\nu}.
 \eqno(17)$$ 
 \par Note that the matrices 
$$ \gamma_n^\mu = \gamma_\lambda \pi^{\lambda \mu}, \eqno(18)$$
 with the projection
$$ \pi^{\lambda \mu}= g^{\lambda\mu} + n^\lambda n^\mu ,\eqno(19)$$
appearing in $(15)$,
 plays an important role in the description of the dynamics in the induced
representation. In $(13)$, the existence of projections on each index
in the spin coupling term implies that $F^{\mu\nu}$ can be replaced by
  ${F_n}^{\mu\nu}$, a tensor projected into the foliation 
subspace. As we shall see, this foliation, induced by the spin, has a
profound effect on the tensor products (and therefore on the full Fock
space) of identical particle systems, both in the boson and fermion
sectors\footnote{\S}{Note that for the $SO(1,1)$ covariant
generalization of one dimensional systems treated, for example, by
methods of the Bethe
ansatz [20], the relation between spin and statistics is not so direct,
and therefore this problem requires a separate discussion.}.

\bigskip
\noindent{\bf 3. The many body problem with spin, and spin-statistics}
\bigskip

\par As in the nonrelativistic quantum theory, one represents the
state of an $N$-body system in terms of a basis given by the tensor
product of $N$ one-particle states, each an element of a one-particle
Hilbert space. The general state of such an
$N$-body system is given by a linear superposition over this
basis [21]. Second quantization then corresponds
to the construction of a Fock space, for which the set of all $N$
body states, for all $N$,  are imbedded in a large Hilbert space for
which operators that change the number $N$ are defined [2]. We
shall discuss this structure in this section, and show, with our
discussion of the relativistic spin given in the previous section,
that the spin of a relativistic
many-body system can be well-defined and, furthermore, that the
quantum fields associated with the particles of the system carry the
induced foliation structure. 
\par In order to
construct the tensor product space corresponding to the many-body
system, we consider, as for the nonrelativistic theory, the product of
wave functions which are elements of isomorphic Hilbert spaces.  In the
nonrelativistic theory, this corresponds to functions at equal time;
in the relativistic theory, the functions are at equal
$\tau$. Thus, in the relativistic theory, there are correlations at
unequal $t$, within the support of the Stueckelberg wave
functions. Moreover, for particles with spin we argue, as a
consequence of the spin-statistics relation, that in the
induced representation, these functions must be taken at {\it identical
values of $n^\mu$}, i.e., taken at the same point on the orbits of the
induced representation of each particle:
\bigskip
\noindent {\it Statement: Identical particles must be represented in
tensor product states by wave functions not only at equal $\tau$ but
also at equal $n^\mu$. }
\bigskip
\par The proof of this statement lies in the observation that the
spin-statistics relation appears to be a universal fact of nature.  The
elementary proof of this statement, for example, for
a system of two spin $1/2$ particles, is that a $\pi$ rotation of the
system introduces
a phase factor of $e^{i {\pi \over 2}}$ for each particle, thus
introducing a minus sign for the two body state.
However, the $\pi$ rotation is equivalent to an interchange of the two
identical particles.  This argument rests on the fact that each
particle is in the same representation of $SU(2)$, which can only be
achieved in the induced representation with the particles at the
same point on their respective orbits.  We therefore see that
identical particles must carry the same value of $n^\mu$, and the
construction of the $N$-body system must follow this
rule\footnote{\dag}{Note that symmetrization and
antisymmetrization can, of course, be carried out with factors in the
tensor product on any sequence in $n$, but the symmetry properties
would not then correspond to the phases associated with spin.} 
It therefore follows that the two body relativistic system can carry a
spin computed by use of the usual Clebsch-Gordon coefficients, and
entanglement would follow even at unequal time (within the support of
the equal $\tau$ wave functions), as in the proposed experiment of
Palacios {\it et al}[12]. This argument can be followed, as we shall
do in Section 4, for arbitrary $N$, and therefore the Fock space of
the quantum field theory carries
the properties usually associated with fermion (or boson) fields. with
the entire Fock space foliated over the orbit of the inducing vector
$n^\mu$.
. 
\par  Although, due to the Newton-Wigner
problem [10] noted above, the solutions of the Dirac equation are not
suitable for the covariant local description of a quantum theory, the
functions constructed in $(11)$ can form
the basis of a consistent, local (off-shell) covariant quantum theory.
\par To show how the many body Fock space develops, we start by constructing a
two body Hilbert space in the
framework of the relativistic quantum theory. The states of this two
body space are given by linear combinations over the product wave
functions, where the wave functions are given by Dirac functions of
the type described in $(11)$,  {\it i.e.}, temporarily suppressing the
indices $n,\tau$,

$$ \psi_{ij}(x_1,x_2) = \psi_i(x_1) \times \psi_j(x_2), \eqno(20)$$
where $ \psi_i(x_1)$ and $\psi_j(x_2)$ are elements of the
one-particle Hilbert space ${\cal H}$. Let us introduce the notation,
often used in differential geometry, that
$$ \psi_{ij}(x_1,x_2) = \psi_i \otimes \psi_j (x_1, x_2), \eqno(21)$$
identifying the arguments according to a standard ordering. Then,
without specifying the spacetime coordinates, we can write
 $$ \psi_{ij} = \psi_i \otimes \psi_j , \eqno(22)$$
formally, an element of the tensor product space ${\cal H}_1 \otimes
{\cal H}_2$. The scalar product is carried out by pairing the elements
in the two factors according to their order, since it corresponds to
 integrals over $x_1, x_2$, {\it i.e.},
$$(  \psi_{ij},  \psi_{k,\ell}) = (\psi_i, \psi_k)(\psi_j, \psi_\ell).
\eqno(23) $$
\par For two identical particle states satisfying
Bose-Einstein of Fermi-Dirac statistics, we must write, according to
our argument given above,
$$ \psi_{ijn}= {1 \over \sqrt 2}[\psi_{in} \otimes \psi_{jn} \pm
\psi_{jn} \otimes \psi_{in}]. \eqno(24)$$
 This expression has the required symmetry
or antisymmetry only if both functions are on the same points of their
respective orbits in the induced representation. Furthermore, they
transform under the {\it same}  $SU(2)$ representation of the rotation
subgroup of the Lorentz group, and thus for spin $1/2$ particles,
under a $\pi$ spatial rotation (defined by the space orthogonal to the timelike
vector $n^\mu$)
they both develop a phase factor $e^{i{\pi \over 2}}$.  The product
results in an over all negative sign.  As in the usual quantum theory,
this rotation corresponds to an interchange of the two particles, but
here with respect to a ``spatial'' rotation around the vector $n^\mu$. The
spacetime coordinates in the functions are rotated in this (foliated)
subspace of spacetime, and correspond to an actual exchange of the
positions of the particles in space time, as in the formulation of
the standard spin-statistics theorem. It therefore
follows that the interchange of the particles occurs in the foliated
space defined by $n^\mu$. For identical bosonic particles, the $\pi$
rotation produces a positive sign.  These conclusions are valid for
unequal times that lie in support of the SHP wave functions (at equal
$\tau$).  We therefore have the 
\bigskip 
\noindent {\it Statement:  The antisymmetry of identical half-integer
spin (fermionic)
particles remains at unequal times (within the support of the wave
functions). This is true for the symmetry of identical integer spin
(bosonic) particles as well}.
\bigskip
\par Furthermore, the construction we have given enables us to define
the spin of a many body system, even if the particles are relativistc
and moving arbitrarily with respect to each other.  Since all
particles with representations on a common $n^\mu$ of their orbits
transform in the spacelike submanifold orthogonal to $n^\mu$ under the
same $SU(2)$, it is also true that  
\bigskip 
\noindent{\it Statement: The spin of an $N$-body system of identical
paticles is well-defined, independent of
the state of motion of the particles of the system, by the usual laws
of combining representations of $SU(2)$, i.e, with the usual
Cebsch-Gordan coefficients, since the states of all the particles in the
system are in induced representations at the same point of the orbit
$n^\mu$.}
\bigskip
\par Thus, for example, in the quark model for hadrons, the total spin of the
hadron can be computed from the spins (and orbital angular momenta
projected into the foliated space) of the individual quarks using
the usual Clebsch-Gordan coefficients even if they are in significant
relative motion, within
the same $SU(2)$; a similar conclusion would be valid for nucleons in
a nucleus even at high excitation. The validity of spin assignations
in high energy scattering would provide an important example of such
quantum mechanical correlations.
  \par In the course of our construction, we have seen that the
foliation of the spacetime follows from the arguments based in the
representations of a relativistic particle with half-integer spin.
However, as we have remarked, our
considerations of the nature of identical particles, and their
association with the spin statistics properties observed in nature,
require that the foliation persists in the bosonic sector as well,
where a $\pi$ rotation, exchanging two
particles, must be in a definite representation of the rotation group,
specified by the foliation vector $n^\mu$, to achieve a positive sign.
Since there is no extra phase
(corresponding to integer representations of the $SU(2)$) for the
 Bose-Einstein case, the boson symmetry
can then be extended to a covariant symmetry with important
implications, for example, for the statistical mechanics of
relativistic boson systems, as, for example, 
Bose-Einstein condensation.  
\par We remark in this connection
that the Cooper pairing[22] of superconductivity must be between electrons
on the same point of their induced representation orbits, so that the
superconducting state is defined on the corresponding foliation of
spacetime as well. The resulting (quasi-) bosons have the identical
particle properties inferred from our discussion of the boson sector.
As remarked above, the two electrons of the Cooper pair may not be at
equal time, a result which may be accessible to experiment.  A similar
remark applies to the Josephson effect[23] (where a single gate may be
opened at two successive times, as in the Lindner experiment[24]).
\par These results have, moreover, important implications in atomic
and molecular
physics, for example,
for the construction of the exchange interaction.
\bigskip
\noindent{\bf 4. Quantum Fields}
 \bigskip
\par We now extend our argument for the finite Fock space to
the general structure of quantum field theory.  
\par The $N$ body state of Fermi-Dirac particles can
 be written as (the $N$ body boson system should be treated
separately since the normalization conditions are different, but we
give the general result below)
$$\Psi_{nN} = {1 \over N!} \Sigma (-)^P P \psi_{nN}\otimes \psi_{nN-1}\otimes
\cdots \psi_{n1}, \eqno(25)$$
where the permutations $P$ are taken over all possibilities, and no
two functions are equal. By the arguments given above, any pair of
particle wave functions in this set have the Fermi-Dirac symmetry properties.
We may now think of such a function as an element of a larger Hilbert
space, the {\it Fock space}, which contains all values of the number
$N$. On this space, one can define an operator that adds another
particle (in the tensor product), performs the necessary
antisymmetrization, and changes the normalization appropriately.  This
operator is called a {\it creation operator}, which we shall denote by
$a^\dagger(\psi_{nN+1})$ and has the property that
$$ a^\dagger(\psi_{nN+1})\Psi_{nN}= \Psi_{nN+1} , \eqno(26)$$
now to be evaluated on the  manifold $(x_{N+1}, x_N, x_{N-1}\dots
x_1)$. Taking the scalar product with some $N+1$ particle state
$\Phi_{nN+1}$ in the Fock space, we see that
$$(\Phi_{nN+1},a^\dagger(\psi_{nN+1})\Psi_{nN}) \equiv
(a(\psi_{nN+1})\Phi_{nN+1}, \Psi_{nN}), \eqno(27)$$
thus defining the {\it annihilation} operator
$a(\psi_{nN+1})$.
\par The existence of such an annihilation operator, as in the usual
construction of the Fock space, {\it e.g.}, [2], implies the existence of
an additional element in the Fock space, the {\it vacuum}, or the state
of no particles.  The vacuum defined in this way lies in the foliation
labelled by $n^\mu$. The covariance of the construction, however,
implies that, since all sectors labelled by $n^\mu$  are connected by
the action of the Lorentz group, that this vacuum is a 
vacuum for any $n^\mu$, {\it i.e.}, the vacuum $\{\Psi_{n0}\}$ over all
$n^\mu$ is Lorentz invariant. 
\par The commutation relations of the annihilation- creation
operators can be easily deduced from a low dimensional example,
following the method used in the nonrelativistic quantum
theory[2]. Consider the two body state $(24)$ (we use the
antisymmetric form here), and apply the creation
operator $a^\dagger (\psi_{n3})$ to create the three body state
$$\eqalign{\Psi(\psi_{n3},\psi_{n2},\psi_{n1})&=
{1 \over \sqrt{3!}}\{\psi_{n3}\otimes\psi_{n2}\otimes \psi_{n1}
+\psi_{n1}\otimes\psi_{n3}\otimes
\psi_{n2} \cr &+ \psi_{n2}\otimes\psi_{n1}\otimes 
\psi_{n3}-\psi_{n2}\otimes\psi_{n3}\otimes
\psi_{n1}\cr &-\psi_{n1}\otimes\psi_{n2}
\otimes \psi_{n3}-\psi_{n3}\otimes\psi_{n1}\otimes
\psi_{n2}\}\cr} \eqno(28)$$
One then takes the scalar product with the three body state
$$\eqalign{\Phi(\phi_{n3},\phi_{n2},\phi_{n1})&=
{1 \over \sqrt{3!}}\{\phi_{n3}\otimes\phi_{n2}\otimes \phi_{n1}
+\phi_{n1}\otimes\phi_{n3}\otimes
\phi_{n2} \cr &+ \phi_{n2}\otimes\phi_{n1}\otimes 
\phi_{n3}-\phi_{n2}\otimes\phi_{n3}\otimes
\phi_{n1}\cr &-\phi_{n1}\otimes\phi_{n2}
\otimes \phi_{n3}-\phi_{n3}\otimes\phi_{n1}\otimes
\phi_{n2}\}\cr} \eqno(29)$$
Carrying out the scalar product term by term, and and picking out the
terms corresponding to the scalar product of some function with the two
body state 
$$ { 1\over \sqrt{2}}\{\psi_{n2}\otimes \psi_{n1}- \psi_{n1}\otimes
\psi_{n2}\} \eqno(30)$$
one finds that the action of the operator $a(\psi_{n3})$ on
the state $\Phi(\phi_{n3},\phi_{n2},\phi_{n1})$ is given by
$$\eqalign{a(\psi_{n3})\Phi(\phi_{n3},\phi_{n2},\phi_{n1})&=
(\psi_{n3},\phi_{n3})\phi_{n2}\otimes\phi_{n1}\cr &-
(\psi_{n3},\phi_{n2})\phi_{n3}\otimes\phi_{n1}
+(\psi_{n3},\phi_{n1})\phi_{n3}\otimes\phi_{n2},\cr} \eqno(31)$$
{\it i.e.}, the annihilation operator acts like a derivation with
alternating signs due to its fermionic nature; the relation of the two
and three body states we have analyzed has a direct extension to the
$N$-body case. The action of boson
annihilation-creation operators can be derived in a similar way.
\par Applying these operators to $N$ and $N+1$ particle states, one
find directly their commutation and anticommutation relations
$$ [a(\psi_n), a^\dagger(\phi_n)]_{\mp} = (\psi_n,\phi_n), \eqno(32)$$
where the $\mp$ sign, corresponds to commutator or anticommutator
for the boson or fermion operators. If the functions $\psi_n,\phi_n$
belong to a normalized orthogonal set $\{\phi_{nj}\}$, then
$$[a(\phi_{ni}, a^\dagger(\phi_{nj})]_{\mp} = \delta_{ij}, \eqno(33)$$
Let us now suppose that the functions $\phi_{nj}$ are plane waves in
spacetime, {\it i.e.}, in terms of functions
$$ \phi_{np}(x)= { 1 \over (2\pi)^2}e^{-ip^\mu x_\mu},\eqno(34)$$
so that
$$ (\phi_{np}, \phi_{np'}) = \delta^4 (p-p'). \eqno(35)$$
The quantum fields are then constructed as follows. Define
$$ \phi_n(x) \equiv \int d^4 p a(\phi_{np})e^{ip^\mu x_\mu}. \eqno(36)$$
It then follows that, by the commutation (anticommutation) relations
$(32)$, these operators obey the relations
$$ [ \phi_n(x), \phi_n^\dagger(x')]_{\mp} = \delta^4 (x-x'), \eqno(37)$$
corresponding to the commutation relations of bose and fermion
{\it fields}(we suppress the spinor indices here, arising from the
spinor form which must be used for $(34)$). Under Fourier transform, 
one finds the commutation relations in momentum space
$$[ \phi_n(p), \phi_n^\dagger(p')]_{\mp} = \delta^4 (p-p')\eqno(38)$$
The relation of these quantized fields with those of the usual
on-shell quantum field theories can be understood as follows. Let us
suppose that the fourth component of the energy-momentum is $E=
\sqrt{{\bf p}^2 + m^2}$, where $m^2$ is close to a given number, the
on-shell mass of a particle. Then, noting that $dE = {dm^2 \over 2E}$,
if we multiply both sides of $(38)$ by $dE$ and integrate over the
small neighborhood of $m^2$ occurring in both $E$ and $E'$, the delta
function $\delta (E-E')$ on the right hand side integrates to unity.
On the left hand
side, there is a factor of $dm^2/2E$, and we may absorb $\sqrt{dm^2}$ in
each of the field variables, obtaining (with $\varphi_n({\bf p}) \equiv
\sqrt{dm^2}\phi_n(p) $ on shell)
$$ [\varphi_n({\bf p}), \varphi_n^\dagger({\bf p'})]_{\mp} =
 2E\delta^3 ({\bf p} -{\bf p'}), \eqno(39)$$
the usual formula for on-shell quantum fields.
\par We remark that these
algebraic results have been constructed in the foliation involved in
the formulation of a consistent theory of relativistic spin, therefore
admitting the action of the $SU(2)$ group (in the Dirac representation
$(12)$, $S(\Lambda)$ has the form $e^{i\Sigma^{\mu\nu}_n
 \omega_{\mu\nu}}$ for $ \omega_{\mu\nu}$ parameters corresponding to
 the $SU(2)$ subgroup leaving $n^\mu$ invariant) for 
a many body system, applicable for unequal times, within the support
of the Stueckelberg wave functions at equal $\tau$. 
\par We have discussed here the construction of quantum fields as they
emerge from the structure of a Fock space. Local observables can be
formed from the Hermitian operators built with these fields.
According to the methods generally attributed to Schwinger[25] and
Tomanaga[26] (see the book of Jauch and Rohrlich[27], for example, for a
discussion of the ideas and additional references), a quantum state
is defined by assigning values to the spectra of a complete set of
such local observables which necessarily commute, according to the
causal nature of measurements, if they are associated
with a spacelike surface. The sequence of spacelike surfaces then
forms a parametrization of  the evolution of such states \footnote{\S\S}{In a 
variational sense.}(the basis of the
Schwinger-Tomonaga equation);
it follows from our considerations that,
for states of identical particles, the set of local observables are
defined on the foliation provided by the inducing parameter $n^\mu$,
and therefore the Schwinger-Tomonaga state lies on this foliation as well.  
Furthermore, since the local fields, in Heisenberg picture, evolve
unitarily in $\tau$, and the corresponding spacelike surfaces are
isomorphic, a correspondence can be established between $\tau$ and an
invariant parameter labelling the sequence of spacelike
surfaces. Moreover, it is clear from $(15)$ that the action of the
operators $\Sigma_n^{\mu\nu}$, due to the occurrence of the
projections $g^{\nu\lambda} + n^\nu n^\lambda$ in the coefficients of
the Lie algebra, correspond to rotations in the spacelike surface
orthogonal to the
timelike vector $n^\mu$ (as we have remarked, in the frame for which
$n^\mu = (1,0,0,0)$, these operators reduce to the ordinary Pauli
matrices). Together with the operators $K^\mu$, they constitute a
representation of the Lorentz group, forming the fundamental
representation of a group oriented with its maximal compact subgroup,
corresponding to the $SU(2)$ little group of Wigner, acting on the
wavefunctions, and the corresponding quantum fields,  as a
rotation in the spacelike surface orthogonal to $n^\mu$. We may
therefore identify the spacelike surfaces on which the quantum fields
are defined with the spacelike surfaces on which the little group induces
 rotations (as in the nonrelativistic theory). Local variations in the 
spacelike surfaces, contemplated
by Schwinger and Tomonaga, then correspond as well to local variations
in the orbit of the induced representation, clearly preserving the
local commutation and anticommutation relations\footnote{\dag\dag}{I
am grateful to one of the referees for raising the question of
quantization on a set of spacelike surfaces}. This structure will be
examined in more detail elsewhere.

\bigskip
\noindent{\bf 5. Discussion and Conclusions}
\bigskip
\par We have applied the method of induced representations of 
Wigner (1939)[4] for the description of a relativistic
 particle with spin; adapted to a structure useful in the
framework of a relativistic quantum theory. The method requires that the
representation be induced on an arbitrary timelike vector instead of
the four-momentum used by Wigner. Each point on the orbit of this
timelike vector is associated with an $SU(2)$, its stability
subgroup. Two particles at the same point on their respective orbits
then transform, under rotations in the space orthogonal to the
timelike vector with the same $SU(2)$ and therefore their spins can be
added with the usual Clebsch-Gordan
coefficients. The existence of the relation between spin and statistics in
nature implies that the fermionic antisymmetry between any pair of
identical particles, associated with a $\pi$ rotation of the
two-body subsystem can be valid only for particles on the same points
of their respective orbits. This result introduces a foliation of the
whole Fock space constructed from the many-body tensor product, and
therefore of the corresponding quantum field theory for both bosons
and fermions; we discuss the correspondence of this foliation
 with the structure of quantum field theory defined on a sequence of
 spacelike surfaces[25][26][27]. We shall discuss the consequences for 
fermion and boson
fields, as well as the (foliated) radiation fields generated by their currents,
more fully in a succeeding publication.
\par   One can, in this way, compute the total spin state of a
many-body system, provided all particles are at the same point on
their respective orbits of the timelike inducing vector, as required
for identical particle systems.  Furthermore,
as in the proposed experiment of Palacios {\it et al}[12], the spin
entanglements induced in this way would exist for particles at equal
world time $\tau$, but not restricted, as in standard nonrelativistc
mechanics, to equal time; these correlations should be seen,
according to this theory, for particles at non-equal times within the
suppport of the Stueckelberg wave functions. The analysis of Palacios
{\it et al}[12] assumes coherence in time and uses, as in the analysis of
Lindner {\it et al}[24] of their experiment, time dependent solutions of
the nonrelativistic Schr\"odinger equation. This treatment is not
consistent with the basic foundations of the quantum theory[28], but
is expected to provide, as in the Lindner {\it et al}[24] experiment, a
good approximation under some circumstances[28]. 
\par The correlations implied by the existence of Cooper pairs,
forming the foundation of the theory of superconductivity, existing,
according to the nonrelativistic quantum theory only at equal times,
are predicted by the SHC theory, as a consequence of the work reported
here, to be maintained at unequal times. The theory can therefore be
generalized to be consistent
with relativistic covariance. In a similar way, we predict that the
interference phenomena associated with the Josephson effect[23] would be
maintained if the two gates are open at different times, or with a
single gate opened at two times, with a result similar to that of the
Lindner {\it et al} experiment[24].  Such a result would be a
remarkable generalization of the Josephson effect.
\par  We finally remark that the Boltzmann counting leading to the
relativistic Bose-Einstein and Fermi-Dirac distributions carried out in
[7] corresponds, from the point of view presented
here, to the use of foliated wave functions; for the free
boson gas or fermion gas, there would be no essential difference in
the results, but the distribution functions would appear with nontrivial
foliation. Furthermore, the Wigner functions in terms of which the
quantum transport was computed in ref [8] would be foliated for the
treatment of identical particle systems, as would the density matrix
and the correspsonding BBGKY hierarchy and the resulting Fokker-Planck
equations. The consequences of this foliation for quantum statistical
mechanics will be discussed more completely, as well, in succeeding
publications.
\bigskip{\it Acknowledgements}
\smallskip
\par I wish to thank Andrew Bennett for many discussions on the
content of this work.     
\bigskip
\noindent{\it References}
\frenchspacing
\item{ 1.}See, for example, L.C. Biedenharn and J.D. Louck, {\it The
Racah-Winger Algebra in Quantum Theory} Addison-Wesley, Reading
Massachusetts (1981)
\item{ 2.}G. Baym, {\it Lectures on Quantum Mechanics}, W.A. Benjamin,
New York (1969).
\item{ 3.}A. Einstein, B. Podolsky and N. Rosen, Phys. Rev.  {\bf 41},
1881 (1978).
\item{ 4.}E. Wigner, Ann. Math.  {\bf 40}, 149 (1939).
\item{ 5.}G.W. Mackey, {\it Induced Representations of Groups and
Quantum Mechanics}, W.A. Benjamin, New York (1968).
\item{ 6.} E.C.G. Stueckelberg,  Helv. Phys. Acta {\bf
14}, 322-323,588-594 (1941); {\bf 15}, 23-27 (1942). L.P. Horwitz and 
C.  Piron,  Helv. Phys. Acta {\bf 46}, 316-326 (1973).
\item{ 7.} L.P. Horwitz, W.C. Schieve and  Piron, Ann. Phys.{\bf 137},
306-340 (1981).
\item{ 8.}L.P. Horwitz, S. Shashoua and W.C. Schieve,
Physica A {\bf 161}, 300 (1989).
\item{ 9.} S. Weinberg, {\it The Quantum Theory of Fields}, vol. 1,
Cambridge Univ. Press, Cambridge (1995). 
\item{ 10.}T.D. Newton and E. Wigner, Rev. Mod. PHys. {\bf 21}, 400 (1949).
\item{ 11.} L.P. Horwitz, C. Piron and F. Reuse,
 Helv. Phys. Acta  {\bf 48}, 546 (1975); R. Arshansky and L.P. Horwitz,
 J. Phys. A: Math. Gen. {\bf 15}, L659-662 (1982).
\item{ 12.} A. Palacios, T.N. Rescigno, and  C.W. McCurdy,
Phys. Rev. Lett. {\bf 103}, 253001-1-253001-4 (2009).
\item{ 13.}M.C. Land and L.P.Horwitz, J. Phys. A:Math and Gen {\bf 28},
3289 (1995).
\item{ 14.} See, for example, H. Boerner, {\it Representations of
Groups}, North Holland, Amsterdam (1963).
\item{ 15.} P.A. M. Dirac, {\it The Principles of Quantum
Mechanics},Clarendon Press, Oxford (1948).
\item{ 16.} J.D. Bjorken and S.D. Drell, {\it Relativistic Quantum
Mechanics}, McGraw Hill, New York (1964).
\item{ 17.}  D. Saad, L.P. Horwitz and R.I.  Arshansky, Found. Phys. {\bf
19}, 1125-1149 (1989). See also, I Aharonovich and L.P. Horwitz,
Jour. Math. Phys. {\bf 47}, 122902 (2006); {\bf 51}, 
052903-1-052903-27 (2010); {\bf
52}, 082901-1-08290111 (2011); {\bf 53}, 032902-1-032902-29 (2012); 
Eur. Phys. Lett. {\bf 97}, 60001-p1-60001-p3 (2012). See also,
J.D. Jackson, {\it Classical Electrodynamics}, 2nd edition, Wiley, New
York (1975).
\item{ 18.} J. Schwinger,Phys. Rev. {\bf 82}, 664 (1949).
\item{ 19.} A.F. Bennett, J. Phys. A:Math. Theor. {\bf 45}, 285302 (2012).
\item{ 20.} See V.E. Koperin, N.M. Bogliubov and A.G. Izergin, {\it
Quantum inverse scattering method and correlation functions,}
Cambridge Monographs on Mathematical Physics, Cambridge (1993).
\item{ 21.} For example, A.L. Fetter and J.D. Walecka, {\it Quantum
Theory of Many Particle Systems}, McGraw-Hill, New York (1971).
\item{ 22.} H. Fr\"ohlich, Phys. Rev. {\bf 79}, 845 (1950);
L.N. Cooper, Phys. Rev. 1189 (1956).
\item{ 23.} B.D. Josephson, Phys. Lett.  {\bf 1}, 251 (1962).
\item{ 24.} F. Lindner {\it et al}  Phys. Rev. Lett.  {\bf 95}, 040401
(2005).
\item{25.} J. Schwinger, Phys. Rev. {\bf 74}, 1439 (1948).
\item{26.} S. Tomanaga, Prog. Theor. Phys. {\bf 1}, 27 (1946).
\item{27.} J.M. Jauch and F. Rohrlich. {\it The Theory of Photons and 
Electrons}. 2nd
edition, Springer-Verlag, New York (1976).
\item{ 28.} L. Horwitz,Phy. Lett. A {\bf 355}, 1-6 (2006).

\vfill
\eject
\end